\documentclass[12pt]{article}
\usepackage[backend=bibtex,style=numeric-comp,sorting=none, citestyle = numeric-comp,maxcitenames=2,maxbibnames=100,uniquelist=false, uniquename=false]{biblatex}
\makeatletter
\providecommand{\bib@extsym}{}
\DeclareEntryOption[string]{extsym}{\renewcommand{\bib@extsym}{#1}}
\renewbibmacro*{begentry}{\printtext{\bib@extsym}}
\makeatother

\usepackage{afterpage}
\usepackage{pdflscape}
\usepackage{amsmath,authblk}
\usepackage{graphicx,psfrag,epsf,empheq}
\usepackage{enumerate}
\usepackage{url} 
\usepackage[top=1in, bottom=1in, left=1in, right=1in]{geometry} 
\usepackage{pdflscape}
\usepackage{color,amssymb,amsmath,graphicx,float,booktabs,subfig,amsthm,bm,bbm,multirow,threeparttable,textcomp,adjustbox,setspace,array,amstext,esint}
\usepackage[hidelinks]{hyperref}
\usepackage{amsthm,soul,cancel,mathrsfs,euscript,amsfonts,xcolor,tikz,bm}
\theoremstyle{plain}

\newtheorem{assumption}{Assumption}
\newtheorem*{assumption*}{\assumptionnumber}
\providecommand{\assumptionnumber}{}


\usepackage{pifont}
\newcommand{\tickNo}{\hspace{-1pt}\Large\ding{55}}
\usepackage{pgf,tikz,wrapfig}
\usetikzlibrary{arrows,shapes.arrows,shapes.geometric,shapes.multipart,decorations.pathmorphing,positioning}
\tikzset{
	>=stealth',
	true/.style={
		rectangle,
		draw=black, very thick,
		text width=6.5em,
		minimum height=2em,
		text centered,
		fill=gray, opacity = 0.5},
	punkt/.style={
		rectangle,
		rounded corners,
		draw=black, very thick,
		text width=6.5em,
		minimum height=2em,
		text centered},
	est/.style={
		circle,
		draw=black, very thick,
		text centered},
	shade/.style={
		circle,
		draw=black, very thick, fill=gray!50,
		text centered},
	weight/.style={
		circle,
		draw=black, very thick,
		text width=6.5em,
		minimum height=2em,
		text centered},
	pil/.style={
		->,
		thick,
		shorten <=2pt,
		shorten >=2pt,},
	double/.style={
		<->,
		very thick,
		shorten <=2pt,
		shorten >=2pt,},
	dash/.style={
		dashed,
		very thick,
		shorten <=2pt,
		shorten >=2pt,},
	dashdouble/.style={
		<->,
		dashed,
		thick,
		shorten <=2pt,
		shorten >=2pt}}
\newcommand{\ind}{\perp\!\!\!\perp}
\newcommand{\nind}{\not\!\perp\!\!\!\perp}

\def\addalpha {\beta_{\scriptscriptstyle \text{Y0}}+}
\def\addbeta {\beta_{\scriptscriptstyle \text{W0}}+}
\def\addgamma {\beta_{\scriptscriptstyle \text{U0}}+}

\begin{document}
\title{A Selective Review of Negative Control Methods in Epidemiology}
\author[1]{Xu Shi\thanks{Email: shixu@umich.edu. The authors have no conflicts to disclose. Human and Animal Rights: This article does not contain any studies with human or animal subjects performed by any of the authors.}}
\author[2]{Wang Miao }
\author[3]{Eric Tchetgen Tchetgen}
\affil[1]{Department of Biostatistics, University of Michigan}
\affil[2]{School of Mathematical Sciences, Peking University}
\affil[3]{Statistics Department, The Wharton School, University of Pennsylvania}
\renewcommand\Authands{ and }
\date{}
\maketitle

\begin{abstract}
\textbf{Purpose of Review} Negative controls are a powerful tool to detect and adjust for bias in epidemiological research. This paper introduces negative controls to a broader audience and provides guidance on principled design and causal analysis based on a formal negative control framework.

\textbf{Recent Findings} We review and summarize causal and statistical assumptions, practical strategies, and validation criteria that can be combined with subject matter knowledge to perform negative control analyses. We also review existing statistical methodologies for detection, reduction, and correction of confounding bias, and briefly discuss recent advances towards nonparametric identification of causal effects in a double negative control design.

\textbf{Summary} There is great potential for valid and accurate causal inference leveraging contemporary healthcare data in which negative controls are routinely available.
Design and analysis of observational data leveraging negative controls is an area of growing interest in health and social sciences. Despite these developments, further effort is needed to disseminate these novel methods to ensure they are adopted by practicing epidemiologists.  

\end{abstract}

\vfill


\section{Introduction}
Despite ongoing efforts to improve study design and statistical analysis of epidemiological research, failure to rule out non-causal explanation of empirical findings has prompted substantial discussions in the health science \cite{ioannidis2005most,hernan2016using}. A powerful tool increasingly recognized to mitigate bias is negative control study design and analysis \cite{lipsitch2010negative,arnold2016brief,arnold2016negative}. 
Negative controls have a long history in laboratory experiments and epidemiology \cite{rosenbaum1989role,weiss2002can,lipsitch2010negative,glass2014experimental}. 
However, they have mainly been used to detect bias rather than to remove bias. More recent methodological advances that enable both bias detection and bias removal have not been fully recognized. 
As a result, the potential for valid and accurate causal inference leveraging contemporary healthcare data with abundant negative controls has to date not been fully realized.
 This paper aims to introduce negative controls to a broader audience and provide guidance on principled design and causal analysis based on a formal negative control framework. 
 We focus on resolving bias due to unmeasured confounding in observational studies, although negative controls have recently also been used to tackle a variety of biases such as selection bias \cite{arnold2016brief,lipsitch2010negative,cai2008identifying}, measurement bias \cite{arnold2016brief,lipsitch2010negative}, and homophily bias \cite{liu2020regression,egami2018identification} in both observational studies and randomized trials \cite{arnold2016negative}.

\subsection{Definition and notation
\label{sec:notation}}
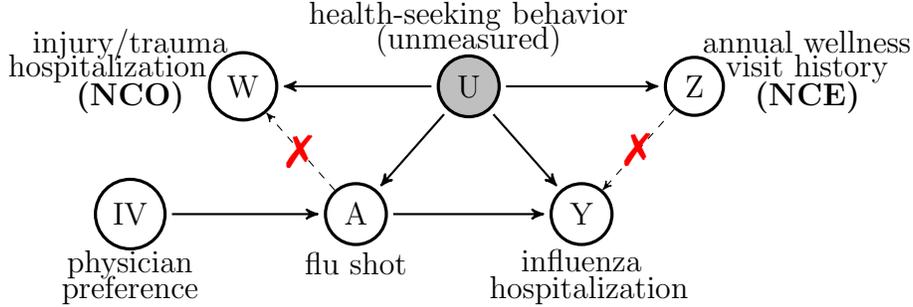
\begin{figure}[h]
	\centering\begin{tikzpicture}
	\node[est]  (A) at (0.5,-1.5) {A};
	\node[est] (Y) at (3.5,-1.5) {Y};
	\node[shade] (U) at (2,0.5-0.3) {U};
	\node[est] (Z) at (5,0.5-0.3) {Z};
	\node[est] (W) at (-1,0.5-0.3) {W};
	\node[est] (IV) at (-2.5,-1.5) {IV};
	\foreach \from/\to in {A/Y,U/A,U/Y,U/W,U/Z,IV/A}\path[->,pil,black] (\from) edgenode {} (\to);
	\coordinate [label={flu shot}] (treatment1) at (0.5,-2.4-0.05);
	\coordinate [label={influenza}] (outcome) at (3.5,-2.4);
	\coordinate [label={hospitalization}] (outcome1) at (3.6,-2.8-0.05);
	\coordinate [label={health-seeking behavior}] (confounders) at (2,0.6+0.5-0.3);
	\coordinate [label={(unmeasured)}] (confounders) at (2,0.35+0.4-0.3);
	\coordinate [label={annual wellness}] (NCE0) at (6.5,0.8-0.3);
	\coordinate [label={visit history}] (NCE1) at (6.5,0.4-0.3);
	\coordinate [label={  \textbf{(NCE)}  }] (NCE) at (6.5,0-0.3);
	\coordinate [label={injury/trauma}] (NCO0) at (-2.5,0.7-0.3);
	\coordinate [label={hospitalization}] (NCO1) at (-2.8,0.4-0.3);
	\coordinate [label={  \textbf{(NCO)}  }] (NCO) at (-2.5,0-0.3);
	\coordinate [label={physician}] (iv1) at (-2.5,-2.4-0.1);
	\coordinate [label={preference}] (iv2) at (-2.5,-2.8-0.05);
	\path[->,dashed]   (A) edge node[midway, inner sep=-1em]{\textcolor{red}{\tickNo}}  (W);
	\path[->,dashed]   (Z) edge node[midway, inner sep=-1em]{\textcolor{red}{\tickNo}}  (Y);
	\end{tikzpicture} 
	\caption{\label{fig:DAG} An illustrating example of different types of negative controls: consider studying the causal effect of flu shot (A) on influenza hospitalization (Y), subject to confounding by unmeasured health-seeking behavior (U). Annual wellness visit history (Z) is an NCE which does not causally affect Y. Injury/trauma hospitalization (W) is an NCO which is not causally affected by A. Both Z and W are proxies of health-seeking behavior. Physician's prescribing preference (IV) is an instrumental variable which likely induces variation in the choice of treatment, and may not affect the outcome other than through its influence on the treatment.  As discussed in Sections~\ref{sec:notation} and \ref{sec:detect}, both a valid instrumental variable and an invalid instrumental variable associated with U are valid NCE.
				All arguments are made implicitly conditional on measured covariates X. Independence between A and Z (or Y and W) conditional on U is not necessary. 
				See more examples in Table~\ref{example_ZW} of the Appendix.}
\end{figure}

A negative control outcome (NCO) is a variable known not to be causally affected by the treatment of interest. Likewise, a negative control exposure (NCE) is a variable known not to causally affect the outcome of interest. To the extent possible, both NCO and NCE should be selected such that they share a common confounding mechanism as the exposure and outcome variables of primary interest, although this is not always necessary \cite{miao2018confounding,sofer2016negative}.
These known-null effects have been used to detect residual confounding bias: 
presence of an association between the NCE and the outcome (or between the NCO and the exposure) constitutes compelling evidence of residual confounding bias, while absence of such association implies no empirical evidence of such bias. 
For example, in a study about the effects of influenza vaccination on influenza hospitalization in the elderly (Figure~\ref{fig:DAG}), injury/trauma hospitalization was considered as an NCO as it can not be causally affected by influenza vaccination, but may be subject to the same confounding mechanism mainly driven by health-seeking behavior \cite{jackson2006evidence}. The authors found that despite efforts to control for confounding, influenza vaccination not only appeared to reduce risk of influenza hospitalization after influenza season (risk ratio 0.82, 95\% CI 0.73--0.92), but also appeared to reduce risk of injury/trauma hospitalization (risk ratio 0.83, 95\% CI 0.75--0.91). This was interpreted as evidence of bias due to inadequately controlled confounding.
Likewise, annual wellness visit history can be considered as an NCE as it is unlikely to cause flu-related hospitalization.

In the following, we adopt the potential outcome framework which we use to formally define causal effects as well as to articulate sufficient identification conditions to perform valid causal inferences from observational data.  We proceed under the fundamental assumption that for each subject in the target population there exist a potential outcome variable $Y(a)$, that would be observed if possibly contrary to fact, the subject were exposed to treatment value $a$, for all possible treatment values of $a$ in a set $\cal{A}$. In the common setting where the treatment is dichotomous $\mathcal{A}=\{0,1\}$, the assumption states that each subject has a well defined pair of potential outcomes $(Y(0),Y(1))$ corresponding to their outcome under active treatment $a=1$ and control treatment  $a=0$, respectively \cite{splawa1990application,rubin1974estimating}. In such setting, our goal is to make inferences about the population average treatment effect (ATE) defined as $\text{ATE}=E[Y(1)-Y(0)]$. 
 Now, consider an observational study in which one observes independent and identically distributed samples on $(Y,A,X)$, where $A$ is a subject's observed binary treatment assignment, $Y$ is his/her observed outcome, and $X$ are observed confounders of the association between $A$ and $Y$. We sometimes refer to $A$ as primary treatment and $Y$ as primary outcome. We assume that the treatment is defined with enough specificity such that among subjects with $A = a$, the observed outcome $Y$ is a realization of the potential outcome value $Y(a)$, that is
 \begin{assumption}[Consistency]\label{assump:consistent}
 	$Y(a)=Y$ when $A=a$.
 \end{assumption}

Much of the literature on causal inference in observational studies  relies on the strong assumption of no unmeasured confounding  for the purpose of identification, i.e., $A\ind Y(a) \mid X$, which is sometimes referred to as ignorability assumption.  This assumption essentially rules out the existence of unmeasured common causes, denoted as  $U$, of the treatment and outcome variables -- an untestable assumption which is often at the source of much skepticism about causal interpretation of associations found in observational data. We do not make such ignorability assumption to establish causation. 
Instead, we invoke the following assumption that describes the relationship between treatment and outcome in the presence of both measured and unmeasured confounding.
\begin{assumption}[Latent ignorability]\label{assump:latent}
	$A\ind Y(a)\mid U,X$.
\end{assumption}
In addition to $(A,Y,X)$, suppose that one has also observed a secondary outcome $W$ and/or a secondary exposure $Z$, and let $Y(a,z)$ and $W(a,z)$ denote the corresponding counterfactual values that would be observed had the primary treatment and secondary exposure taken value $(a,z)$. $W$ and $Z$ are formally defined as negative control outcome and exposure variables provided that the following assumptions hold
\begin{assumption}[Negative control outcome]\label{assump:nco}
	$W(a,z)=W$ and $W\ind A\mid U,X$. 
\end{assumption} 
\begin{assumption}[Negative control exposure]\label{assump:nce}
	$Y(a,z)=Y(a)$ and $Z\ind (Y(a),W)\mid U,X$.
\end{assumption}

Assumptions~\ref{assump:nco} and \ref{assump:nce} entail: (1) there is no remaining unmeasured common cause between $(A,Z)$ and $(Y,W)$ conditional on $(U,X)$;
(2) there is no causal effect of $Z$ on $Y$ conditional on $U$, $A$ and $X$, and there is no causal effect of $A$ and $Z$ on $W$ conditional on $U$ and $X$, which are referred to as the exclusion restrictions. We refer to a pair of $W$ and $Z$ as the double negative control. It is not necessary to have both NCO and NCE, although the double negative control will be sufficient for nonparametric identification of the ATE as detailed in Section~\ref{sec:reducecorrect}.

Figure~\ref{fig:DAG} illustrates a directed acyclic graph (DAG) encoding the above assumptions. Consider a study of the effectiveness of flu shot ($A$) on influenza-related hospitalization ($Y$). A major concern in such studies is potential hidden bias due to unmeasured health-seeking behavior ($U$), a well-known common cause of flu shot status and influenza hospitalization. 
In such a study, routinely captured information on a person's annual wellness visit history entails a good candidate NCE ($Z$) satisfying Assumption~\ref{assump:nce}, as it reflects a person's tendency to engage in healthy behavior, and is unlikely to cause influenza hospitalization. 
Similarly, recorded data on a person's injury/trauma hospitalization provides compelling candidate NCO($W$) satisfying Assumption~\ref{assump:nco}, as it is likely associated with health-seeking behavior and unaffected by flu shot. In addition, we can view an instrumental variable (IV) as an NCE  \cite{miao2018confounding,shi2018multiply}. An IV is a pre-treatment variable satisfying the following three core assumptions: (IV relevance) the IV must be associated with the treatment; (Exclusion restriction) the IV must not have a direct effect on the outcome that is not mediated by the treatment; (IV independence) the IV must be independent of unmeasured confounders. For example, physician’s prescribing preference is often taken as an IV in comparative effectiveness studies, because it likely induces variation in the choice of treatment, and may not affect the outcome other than through its influence on the treatment \cite{brookhart2010instrumental}.  A valid IV satisfies Assumption~\ref{assump:nce} and hence is a valid NCE, which is further explained in Section~\ref{sec:detect}. Besides the above three IV conditions, a forth condition is necessary to identify a causal effect, such as the monotonicity assumption or the no current treatment interaction assumption \cite{angrist1996identification,hernan2006instruments,robins1994correcting,wang2018bounded}.  Alternatively, causal effect identification using IV is also made possible by further incorporating an NCO under a double negative control framework introduced in Section~\ref{sec:reducecorrect}.

It is important to note that Figure~\ref{fig:DAG} is not the only DAG satisfying the negative control assumptions. For example, a more general DAG would allow $Z$ to affect $A$, corresponding to the case where an annual wellness visit could result in flu vaccination during flu season. Moreover, physician preferences are not randomized and may be associated with $U$ via physician-patient interactions, potentially violating the IV independence assumption. Such an invalid IV violating the IV independence assumption is still a valid NCE as long as the exclusion restriction holds, regardless of whether the IV relevance assumption holds. In this case, an NCO can be used to repair an invalid IV for causal effect identification under a double negative control framework \cite{miao2018confounding,shi2018multiply}. 
Additional DAGs illustrating settings in which Assumptions \ref{assump:latent}-\ref{assump:nco} hold are provided in Table~\ref{example_ZW} of the Appendix.
As demonstrated in \cite{miao2018confounding} and \cite{shi2018multiply}, an NCE can be either pre- or post-treatment variable. 
Unmeasured common causes of the $Z$-$A$ association and $Y$-$W$ association can also be present without necessarily invalidating Assumptions~\ref{assump:nco}-\ref{assump:nce}. A key insight is that a valid NCO does not necessarily need to be an outcome variable and may in fact precede the treatment in view, while a valid NCE need not necessarily be a treatment and may in fact be ascertained either together with primary outcome of interest or subsequently.

\subsection{Inconsistent terminology  in literature}
In  prior literature, NCO has been referred to as falsification outcome/end point \cite{prasad2013prespecified, markovitz2019performance, bijlsma2016effect,lin2018global}, control outcome \cite{jackson2006evidence, dusetzina2015control,rosenbaum2010design}, secondary outcome \cite{munafo2018collider,mealli2013using}, supplementary response \cite{rosenbaum1989role} and unaffected outcome \cite{rosenbaum1992detecting}. NCE has been referred to as control exposure \cite{dusetzina2015control} and residual-confounding indicator \cite{ flanders2011method, flanders2017new}. Both NCO and NCE have been referred to as proxies of unmeasured confounder \cite{de2017proxy,kuroki2014measurement,miao2018identifying}. In addition, an exposure-outcome pair known a priori to be unrelated has also been referred to as a negative control pair \cite{madigan2014systematic,schuemie2014interpreting,schuemie2016robust,schuemie2018empirical,schuemie2018improving}. 

The literature reviewed in the current paper is largely limited to papers that use aforementioned nomenclature.  Although \cite{lipsitch2010negative}  and \cite{dusetzina2015control} review negative control literature, to the best of our knowledge, this paper is the first to systematically summarize both formal causal and statistical methodology together with applications of negative controls.
The rest of the paper is organized as follows. Design and validation of negative controls are discussed in Section~\ref{sec:application}. We then review both assumptions and methods for using negative controls to detect, reduce, and remove unmeasured confounding bias in Section~\ref{sec:method}. We use a simple example to illustrate double negative control adjustment (i.e., leveraging NCE and NCO when both are available) of confounding bias in Section~\ref{sec:reducecorrect}. We close with a summary in Section~\ref{sec:discussion}.

\section{Review of applications\label{sec:application}}
Existing applications of negative controls mainly focus on detection of uncontrolled confounding bias.
We list in Table~\ref{tbl:application} selected studies that employed negative controls to detect residual confounding and to strengthen causal conclusions. Among these studies, eight used NCEs and nine used NCOs. Table~\ref{tbl:application} is by no means comprehensive, as hundreds of studies have leveraged negative control variables as evidenced by the number of recent articles that have cited \cite{lipsitch2010negative} as the foundational paper on the use of negative control exposures and outcomes in Epidemiology, but rather a representative set of examples that help illustrate strategies for identifying compelling candidate negative controls.
\subsection{Examples of negative control designs}
\paragraph{Effect of influenza vaccination on influenza hospitalization: using injury/trauma hospitalization as an NCO} As detailed in Section~\ref{sec:notation}, 
to study the effects of influenza vaccination on influenza hospitalization in the elderly, injury/trauma hospitalization was taken as an NCO to detect confounding by unmeasured health-seeking behavior \cite{jackson2006evidence}. Influenza hospitalization before the flu season was also used as an NCO, because flu vaccine can not protect against influenza hospitalization when there is little flu virus circulation.

\paragraph{Effect of maternal exposure on offspring outcomes: using paternal exposure as an NCE} A  number of publications have used paternal exposure as an NCE to study the intrauterine effect of maternal exposure on offspring outcome. Specifically, \cite{yerushalmy1971relationship,mitchell1993smoking,howe2012maternal,brion2007similar,smith2008assessing} studied the association between maternal smoking and offspring outcomes, and compared paternal and maternal associations to detect potential bias due to unmeasured confounding by family-level confounding factors or parental phenotypes. Similarly, \cite{brew2017using} compared maternal and paternal distress and their associations with offspring asthma. 
Evaluation of the validity of paternal exposure as an NCE has also been considered in \cite{taylor2014partner}. They found that cotinine level from exposure to partner smoking were low in non-smoking pregnant women, which suggests that using paternal smoking as an NCE for investigating intrauterine effects is valid.

\paragraph{Effect of air pollution on health outcomes: using future air pollution as an NCE}  Besides use of paternal exposures, NCEs are also used in air pollution studies. For example, \cite{flanders2011method,flanders2017new,miao2017invited,yu2020identification} studied statistical methods that utilize future air pollution as an NCE for bias detection and bias reduction, because the future is not expected to causally affect the past.
In addition, \cite{lumley2000assessing} studied the effect of air pollutant on asthma, and leveraged two different NCEs: air pollutant level in the future and air pollutant level in a distant city.

\afterpage{%
	\clearpage
	\begin{landscape}
		\begin{table}
			\centering{\small
				\begin{tabular}{|>{\raggedright}p{2cm}|>{\raggedright}p{4.5cm}|>{\raggedright}p{4.5cm}|>{\raggedright}p{5.5cm}|>{\raggedright}p{6cm}|}
					\hline 
					\textbf{Reference} & \textbf{Exposure} & \textbf{Outcome} & \textbf{Negative Control Exposure} & \textbf{Negative Control Outcome}\tabularnewline
					\hline 
					\hline 
					\cite{yerushalmy1971relationship} & Maternal smoking & Low birth weight & Paternal smoking & \tabularnewline
					\hline 
					\cite{mitchell1993smoking} & Maternal smoking & Sudden infant death syndrom & Paternal smoking & \tabularnewline
					\hline 
					\cite{howe2012maternal} & Maternal smoking & Offspring height, ponderal index, body mass index & Paternal smoking & \tabularnewline
					\hline 
					\cite{brion2007similar} & Maternal smoking & Offspring blood pressure & Paternal smoking & \tabularnewline
					\hline 
					\cite{brew2017using} & Maternal distress & Offspring asthma & Paternal distress & \tabularnewline
					\hline 
					\cite{smith2008assessing,taylor2014partner}:  & Maternal smoking, alcohol use or dietary patterns & Offspring development & Paternal smoking, alcohol use or dietary patterns & \tabularnewline
					\hline 
					\cite{lumley2000assessing} & Air pollutant & Ashma & Future air pollutant, air pollutant elsewhere & \tabularnewline
					\hline 
					\cite{lousdal2020negative} & Mammography-screening participation & Death from breast-cancer & Dental-care participation & Death from causes other than breast cancer and from external causes
					such as accidents, intentional self-harm and assaults\tabularnewline
					\hline 
					\cite{jackson2006evidence} & Influenza vaccination & Mortality and pneumonia/influenza hospitalization &  & Outcome before and after influenza season; injury/trauma hospitalization\tabularnewline
					\hline 
					\cite{sheppard1999effects} & Air pollutant & Asthma hospitalization &  & Appendicitis hospitalization\tabularnewline
					\hline 
					\cite{hammond1954relationship,doll1954mortality,doll1956lung,cornfield1959smoking} & Smoking & Mortality from lung cancer &  & Other causes of death\tabularnewline
					\hline 
					\cite{trichopoulos1983psychological} & Psychological stress post earthquake & Deaths from cardiac events &  & Other causes of death, e.g. cancer\tabularnewline
					\hline 
					\cite{selby1992case,zauber2015impact} & Screening sigmoidoscopy & Mortality from distal colon tumor &  & Mortality from proximal colon tumor (above the reach of the sigmoidoscopy)\tabularnewline
					\hline 
				\end{tabular}
			}
			\caption{\label{tbl:application} Summary of selected applications using negative controls for detection of confounding bias.}
		\end{table}
	\end{landscape}
}

\subsection{Summary of negative control designs} 
In addition to the above examples, various negative control designs are also summarized  in Table~\ref{tbl:application}. Rather than detailing each study in Table~\ref{tbl:application}, we summarize these studies in terms of their respective strategy to identify negative control variables below.
A commonly used strategy to select negative controls  leverages temporal and spacial constraints that essentially guarantee the exclusion restrictions in Assumptions~\ref{assump:nco}-\ref{assump:nce}. 
Temporal ordering leverages the universal truth that the future cannot causally affect the past. For example, as detailed above, \cite{flanders2011method,flanders2017new,miao2017invited,yu2020identification,lumley2000assessing} specify future measurements of air pollution as an NCE to study the effect of current air pollution on health outcomes. Similarly, \cite{smith2008assessing} proposed to look at maternal exposure before and after pregnancy in studying the intrauterine effect of maternal exposure on offspring outcome. An essential prerequisite for this design is that primary outcome does not cause subsequent exposure (at least in the short term), certainly a reasonable assumption in air pollution settings.
Prior information about timing of exposure also sometimes allows one to leave out an essential ingredient \cite{lipsitch2010negative}. For instance, \cite{jackson2006evidence} defined as NCO the number of hospitalizations prior to  influenza season in order to estimate the  effect of influenza vaccination on influenza hospitalization, as little to no flu circulates prior to flu season for influenza vaccination to be protective against. 
Spatial  distancing  has also been considered as an effective means to enforce exclusion restrictions in Assumptions~\ref{assump:nco}-\ref{assump:nce}. For instance, \cite{lumley2000assessing} took air pollutant level in a distant city as an NCE to study the effect of air pollutant on asthma. \cite{selby1992case,zauber2015impact} studied screening sigmoidoscopy and mortality from colon tumor, and selected tumor from proximal colon that is beyond the reach of the sigmoidoscopy as an NCO.  

Another strategy is to select as NCO an outcome analogous to the primary outcome however resulting from  mechanism a priori known to be unrelated to  the primary treatment. As illustration of this approach, consider \cite{jackson2006evidence} which took hospitalization due to injury/trauma as an NCO for the primary outcome, hospitalization due to influenza.  Similarly, to evaluate the effect of air pollution on hospitalization due to asthma, \cite{sheppard1999effects} defined hospitalization due to appendicitis as an NCO.
In addition, several studies routinely use death from other causes as NCO: \cite{hammond1954relationship,doll1954mortality,doll1956lung,cornfield1959smoking} studied the effect of smoking on lung cancer with mortality from other causes as an NCO, \cite{trichopoulos1983psychological} studied the effect of psychological stress on deaths from cardiac events after an earthquake with death from other causes as an NCO, and \cite{lousdal2020negative} selected death from causes other than breast cancer and from external causes such as accidents, intentional self-harm and assaults as NCO to estimate the effect of mammography-screening participation on breast cancer mortality.

\subsection{Validation of negative controls by subject matter knowledge}
Despite the various strategies in the literature to find candidate negative controls,  researchers should rigorously validate the choice of negative controls and be aware of possible violations of  negative control assumptions.
Similar to the assumptions of no unmeasured confounding,  negative control assumptions (Assumptions~\ref{assump:nco} and \ref{assump:nce}) are causal assumptions that can only be established by subject matter considerations and not by empirical test without additional assumptions. 
In practice, we recommend checking the following criteria in finding a candidate negative control.
\begin{itemize}
	\item ``Irrelevant to $Y$ (or $A$)": The NCE should not cause the outcome of interest, while the NCO should not be caused by the treatment of interest nor the NCE. These conditions are formally implied by Assumptions~\ref{assump:nco} and \ref{assump:nce}.
	\item ``Comparable to $A$ (or $Y$)":  {In most cases it is important to have the source of bias in mind before designing a negative control study although this is not always necessary \cite{miao2018confounding,sofer2016negative}.}  Unmeasured confounding mechanism of negative controls should be comparable to that of $A$ and $Y$ in the following sense: the NCE must be associated with unmeasured confounders conditional on measured confounders and primary treatment; the NCO must be associated with unmeasured confounders conditional on measured confounders. {Hence the negative control variable is often viewed as a proxy of the unmeasured confounders. A variable completely irrelevant to all mechanisms under consideration would not provide any useful information. }  These conditions are formally required by  Assumptions~\ref{assump:ucompare} and \ref{assump:completeness} in Section~\ref{sec:method};
	\item ``Adequate Negative Control Power": The NCE and NCO are not exceedingly rare relative to primary treatment and outcome variables, respectively. For example, in the event that the negative control variable is a rare binary variable, or if the association between unmeasured confounder and negative control variable is  weak, then large sample may be necessary  to achieve sufficient power for detecting confounding bias \cite{hengelbrock2017re,richardson2017negative}.
\end{itemize}
We list examples of possible violations of negative control assumptions in the Appendix.

\section{Review of methods\label{sec:method}}
\subsection{Bias detection\label{sec:detect}}
\paragraph{Key assumption and rationale for bias detection} 
Assumptions \ref{assump:nco} and \ref{assump:nce} give rise to formal statistical tests of the null hypothesis that adjustment for observed covariates suffices to control for confounding bias, rejection of which indicates presence of an unmeasured confounder $U$. A key assumption for this bias detection strategy is that the negative control exposure or outcome is $U$-comparable to the primary exposure or outcome:
\begin{assumption}[$U$-comparable]\label{assump:ucompare}
	$W\nind U\mid X$ and $Z\nind U\mid A,X$.
\end{assumption}
The $U$-comparability assumption requires that unmeasured confounders $U$ of $A$-$Y$ association are identical to those of the $A$-$W$ association and $Z$-$Y$ association, such that a non-null $A$-$W$ or $Z$-$Y$ association can be attributed to $U$. Therefore, presence of an association between primary and  negative control variables implies residual confounding bias, while absence of such associations implies no empirical evidence of unmeasured confounding. 
It is important to note that when evaluating $Z$-$Y$ association one must also adjust for $A$ to rule out the potential association between $Z$ and $Y$ due to the pathway $Z-A\rightarrow Y$  (the arrow between $Z$ and $A$ could either be $Z\rightarrow A$ or $Z\leftarrow A$). Examples of such relationships are listed in Table~\ref{example_ZW} of the Appendix.
Notably, conditional on $X$, a valid IV independent of $U$ and associated with $A$ satisfies Assumption~\ref{assump:ucompare} because of conditioning on a collider $A$ on the $\text{IV}\rightarrow A \leftarrow U$ pathway \cite{shi2018multiply,miao2018confounding}; likewise an invalid IV that violates the IV independence assumption defined in Section~\ref{sec:notation}  would also satisfy Assumption~\ref{assump:ucompare} regardless of whether IV and $A$ are associated, as mentioned in Section~\ref{sec:notation}.
\paragraph{Methods} As detailed in Section~\ref{sec:application}, majority of existing applications used negative controls for bias detection, by testing for an  association between primary and negative control variables. A review of bias detection methods is presented in Table~\ref{tbl:method}. For example,  \cite{flanders2011method} formalized bias detection as a Wald test of the coefficient of NCE in a regression model of the outcome on the primary and negative control exposures. Moreover, \cite{smith2012negative,weisskopf2016commentary} noted that an invalid NCE that violates the exclusion restriction but satisfies the $U$-comparable assumption can nevertheless validate a causal interpretation when it does not appear to be associated with the outcome adjusting for the treatment of interest.

\subsection{Bias reduction and bias correction\label{sec:reducecorrect}}

\paragraph{Summary of literature} Beyond bias detection, recent developments have made it possible to  reduce and sometimes completely remove unmeasured confounding bias using negative controls.
In air pollution studies, current and future pollutant levels are often positively correlated and are associated with  unmeasured confounders in the same direction. In this setting,  \cite{flanders2017new} showed that incorporating future air pollution, an NCE, in the outcome model can reduce confounding bias. Further bias attenuation was proposed in \cite{miao2017invited} by incorporating both past and future exposures. Bias reduction using an NCO was considered by \cite{richardson2015negative} in estimation of standardized mortality ratio, where the standardized mortality ratio of the NCO was used to reduce bias in that of the primary outcome. In addition, \cite{schuemie2014interpreting,schuemie2018empirical} considered 

{\afterpage{%
		\newgeometry{margin=1.1cm} 
		\begin{landscape}
			\begin{table}
				\centering{\small
					\begin{tabular}{|c|>{\raggedright}p{5.1cm}|>{\raggedright}p{9.9cm}|>{\raggedright}p{9cm}|}
						\hline 
						& \textbf{Reference and Setting} & \textbf{Main Assumptions Besides Assumptions~\ref{assump:latent}-\ref{assump:ucompare}} & \textbf{Methods}\tabularnewline
						\hline 
						\hline 
						\multirow{2}{*}{\textbf{D}} & \cite{flanders2011method}: Time-series study.
						
						$Z$ = future air pollution $A_{t+1}$. & (1) $A_{t+1}\perp Y_{t}\mid A_{t},U_{t},X_{t}$.\\
						(2) $\log[E(Y_{t})]=\alpha+\beta A_{t}+\gamma X_{t}+\beta_{f}A_{t+1}$. & Bias detection by Wald-test on $\beta_{f}$.\tabularnewline
						\cline{2-4} 
						& \cite{weisskopf2016commentary,smith2012negative}:
						invalid NCE $Z$. & (1) Violation of exclusion restriction $Y(a,z)\neq Y(a)$.\\
						(2) $Z$ is $U$-comparable with $A$: $Z\nind U\mid A,X$. & No evidence of $Z$-$Y$ association adjusting for $A$ implies no
						residual confounding of $A$-$Y$ association.\tabularnewline
						\hline 
						\multirow{3}{*}{\textbf{R}} & \cite{flanders2017new,miao2017invited}: Time-series
						study.
						
						$Z$ = future air pollution $A_{t+1}$. & (1) $A_{t+1}\perp Y_{t}\mid A_{t},U_{t},X_{t}$; $A_{t+1}\nind(A_{t},U_{t})\mid X_{t}$.\\
						(2) $Y_{t}(a_{t},x_{t},u_{t})=\beta_{0}+\beta_{1}\alpha_{t}+\beta_{2}x_{t}+\beta_{3}u_{t}+\epsilon_{t}$;
						$E[\epsilon_{t}\mid A_{t}=a_{t},U_{t}=u_{t},X_{t}=x_{t}]=0$.
						
						(3) $E[U_{t}\mid A_{t}=a_{t},A_{t+1}=a_{t+1},X_{t=x_{t}}]=\alpha_{0}+\alpha_{1}a_{t}+\alpha_{2}x_{t}+\alpha_{3}a_{t+1};sign(\alpha_{1})=sign(\alpha_{3})$.
						
						(4) $E[A_{t+1}\mid A_{t}\!=\!a_{t},X_{t}\!=\!x_{t}]\!=\!\gamma_{0}\!+\!\gamma_{1}a_{t}\!+\!\gamma_{2}x_{t};\gamma_{1}>0$. & Bias reduction by fitting $E[Y_{t}\mid A_{t},X_{t},A_{t+1}]$ instead
						of fitting $E[Y_{t}\mid A_{t},X_{t}]$. Further bias reduction considered
						in \cite{miao2017invited} by incorporating $X_{t+1}$
						or $A_{t-1}$. Identification of $\beta_{1}$ is possible with multiple
						future exposures under autoregressive model for exposure time series.\tabularnewline
						\cline{2-4} 
						& \cite{richardson2015negative}: Standardized mortality
						ratio in occupational cohort study. & (1) $E[Y(1)\mid X=k]/E[Y_{\text{ref}}\mid X=k]=\exp(\alpha_{k}-\delta_{k})$
						
						$E[W\mid X=k]/E[W_{\text{ref}}\mid X=k]=\exp(-\epsilon_{k})$.
						
						(2) $sign(\epsilon_{k})=sign(\delta_{k})$ and $0<|\epsilon_{k}|<2|\delta_{k}|$. & Adjust for bias $\delta_{k}$ via $E[Y(1)\mid X=k]E[W_{\text{ref}}\mid X=k]/{E[Y_{\text{ref}}\mid X=k]E[W\mid X=k]}.$\tabularnewline
						\cline{2-4} 
						& \cite{schuemie2014interpreting,schuemie2018empirical}:
						Define negative controls as drug--outcome pairs where one believes
						no causal effect exists. & (1) For a negative control drug-outcome pair, the effect estimate
						$\beta_{i}\sim N(\theta_{i},\tau_{i}^{2}),i=1,\ldots,n$, where $\theta_{i}\sim N(\mu,\sigma^{2})$
						is the true bias.
						
						(2) Under the null of no treatment effect, the effect estimate $\beta_{n+1}\stackrel{H_{0}}{\sim}N(\mu,\sigma^{2}+\tau_{n+1}^{2})$. & Estimate $\mu,\sigma$ by MLE with $L(\mu,\sigma\mid\theta,\tau)=\Pi_{i=1}^{n}\int p(\beta_{i}\mid\theta_{i},\tau_{i})p(\theta_{i}\mid\mu,\sigma)d\theta_{i}$.
						Calibrated $p$-value computed via Wald-test of $\beta_{n+1}$. Confidence
						interval calibrated similarly using distribution generated by positive
						controls.\tabularnewline
						\hline 
						\multirow{5}{*}{\textbf{C}} & \cite{richardson2014assessment,tchetgen2015negative}:
						$W,Y$= Time-to-event outcome. & (1) There exist monotonic functions that describe $U$-$Y$ and $U$-$W$
						associations: $Y(0)=h_{y}(U,X),W=h_{w}(U,X)$.
						
						(2) Cox models for $Y$ and $W$ w/ hazard ratio $e^{\beta_{y}}$
						and $e^{\beta_{w}}$. & The hazard ratio measuring the causal effect of treatment is $e^{\beta_{y}-\beta_{w}}$.\tabularnewline
						\cline{2-4} 
						& \cite{sofer2016negative,glynn2019generalized}: Generalized difference-in-differences
						using NCO. & (1) There exist monotonic functions that describe $U$-$Y$ and $U$-$W$
						associations: $Y(0)=h_{y}(U,X),W=h_{w}(U,X)$.
						
						(2) Positivity: if $0<f_{W\mid A=1,X}(W^{*})$ then $0\!<\!f_{W\mid A=0,X}(W^{*})\!<\!1$,
						where $W^{*}=(W\mid A=1,X)$ is distributed as $W$ in the exposed
						group. & The average treatment effect on the treated is $E[Y(1)-Y(0)\mid A=1]=E[Y\mid A=1]-$$E[F_{Y\mid A=0,X}^{-1})\cdot F_{W\mid A=0,X}(W^{*})]$.
						Generalized the difference-in-differences approach to the broader
						context of NCO.\tabularnewline
						\cline{2-4} 
						& \cite{tchetgen2014control}: Calibration using NCO. & (1) $W\perp A\mid X,Y(1),Y(0)$. (2) Rank preservation: $Y=Y(0)+\Psi A$,
						and hence $W\perp A\mid X,Y(0)$ by (1).
						
						(3) $E[W\!\mid\!A,Y(0)\!=\!Y\!-\!\Psi A,X]\!=\!\beta_{1}\!+\!\beta_{2}X\!+\!\beta_{3}Y(\Psi)\!+\!\beta_{4}A$,
						where $\beta_{4}=0$ by (1). & The 95\% CI for any $\Psi_{0}$ consists of all $\Psi$ for which
						$\hat{\beta}_{4}(\Psi)\pm1.96\text{{s.e.}}[\hat{\beta}_{4}(\Psi)]$
						contains 0; Under (1)-(3), fit $E[W\mid A,Y,X]=\beta_{1}+\beta_{2}X+\beta_{3}Y+\beta_{\Psi}A$,
						then the causal effect $\Psi=-\beta_{\Psi}/\beta_{3}$.\tabularnewline
						\cline{2-4} 
						& \cite{gagnon2012using,jacob2016correcting,wang2017confounder}:
						Removing unwanted variation in gene-expression analysis. & (1) $Y_{1\times p}=X_{1\times q}\beta_{q\times p}+U_{1\times r}\Gamma_{r\times p}+\epsilon_{1\times p}$,
						$p\geq r+1$.
						
						(2) $W_{1\times s}=U_{1\times r}\Gamma_{r\times s}^{W}+\epsilon_{1\times s}^{W}$,
						$s\geq r,\text{Rank}(\Gamma_{r\times s}^{W})=r$.
						
						(3) $(\epsilon,\epsilon^{W})\sim N(0,\text{diag}(\sigma_{1}^{2},\dots,\sigma_{p+s}^{2})),(\epsilon,\epsilon^{W})\ind(X,U)$.
						
						(4) $U_{1\times r}=X_{q}\alpha_{q\times r}+\epsilon_{1\times r}^{U}$,
						$\epsilon^{U}\sim N(0,I_{r}),\epsilon^{U}\ind X$. & \cite{gagnon2012using,jacob2016correcting}:
						Estimate $U$ by factor analysis of (2), then estimate $\beta$ from
						(1). \cite{wang2017confounder}: Estimate $\Gamma^{W}$
						and $\Gamma$ by factor analysis of $Y=X(\beta+\alpha\Gamma)+(\epsilon^{U}\Gamma+\epsilon)$
						(5) and $W=X\alpha\Gamma^{W}+(\epsilon^{U}\Gamma^{W}+\epsilon^{W})$
						(6). Then estimate $\alpha$ from (6), and estimate $\beta$ from
						(5).\tabularnewline
						\cline{2-4} 
						& \cite{miao2018identifying,miao2018confounding,shi2018multiply}:
						Nonparametric identification. & 
						Assumption~\ref{assump:completeness}
						& Identify $h$ in $E[Y\!\mid\! A,Z,X]=$$E[h(W,A,X)\!\mid\!A,Z,X]$,
						then $\text{ATE}=E[h(W,A=1,X)]-E[h(W,A=0,X)]$.\tabularnewline
						\hline 
					\end{tabular}
				}
				\caption{\label{tbl:method} Summary of published methodologies using negative controls for detection (D), reduction (R), and correction (C) of confounding bias.}
			\end{table}
		\end{landscape}
		\restoregeometry
}}
\restoregeometry

\clearpage
\noindent calibrating  $p$-value and confidence intervals by deriving an empirical null distribution from the association between primary and negative control variables.

Several methods were developed to achieve full bias removal, under certain assumptions such as monotonicity \cite{richardson2014assessment,tchetgen2015negative,sofer2016negative,glynn2019generalized}, rank preservation \cite{tchetgen2014control}, and linear model for unmeasured confounding. Specifically, \cite{richardson2014assessment,tchetgen2015negative} considered bias correction by using a negative control time-to-event outcome under a monotonicity assumption that describes the $U$-$Y$ and $U$-$W$ association. Under a similar monotonicity assumption, \cite{sofer2016negative} generalized difference-in-difference method to NCO method, which is further extended by \cite{glynn2019generalized}. In addition, \cite{tchetgen2014control} developed an outcome calibration approach with a rank preservation assumption under which the counterfactual primary outcome can account for the unmeasured confounding between the $A$-$W$ association. Lastly, \cite{gagnon2012using,jacob2016correcting,wang2017confounder} assumed a linear model for the unmeasured confounder and proposed to estimate $U$ by factor analysis.

\paragraph{Nonparametric identification in a double negative control design} The above methods remove unmeasured confounding bias under relatively stringent assumptions. \cite{miao2018identifying} established sufficient conditions under which the ATE can be nonparametrically identified leveraging an NCE and an NCO, i.e., via a double negative control design \cite{shi2018multiply}. That is, the ATE can be uniquely expressed as a function of the observed data distribution without imposing any restriction on the observed data distribution, such that distinct data generating mechanisms are guaranteed to lead to distinct ATE values. Further method developments include semiparametric estimation under categorical negative controls and unmeasured confounding  \cite{shi2018multiply} and alternative strategies to identify the ATE via a so-called confounding bridge function \cite{miao2018confounding}.

Double negative controls are widely available in health sciences. For example, in air pollution studies, \cite{miao2018confounding} used future air pollution level and past health outcome as negative control exposure and outcome, respectively. \cite{shi2018multiply} took two routinely monitored control outcomes from administrative healthcare data in vaccine safety studies as double negative control, in the setting where both control outcomes are independent of the primary outcome and satisfy both Assumption~\ref{assump:nco} and Assumption~\ref{assump:nce}. In influenza vaccine effectiveness research presented in Figure~\ref{fig:DAG}, annual wellness visit and injury/trauma hospitalization can serve as double negative control.   In addition, when IV is available, identification is made possible by further incorporating an NCO such as a pretreatment measurement of the outcome.

Below we will first detail the identification conditions established in \cite{miao2018identifying}  and then introduce  identification methods proposed in \cite{miao2018identifying} and \cite{miao2018confounding}.
\begin{assumption}[Positivity]\label{assump:positivity}
	$0<P(A=a,Z=z\mid X)<1$ for all $a$, $z$.
\end{assumption}

\begin{assumption}[Completeness]\label{assump:completeness}
	(a) For all $a$, $W \nind Z\mid A=a,X$. (b) For any square integrable function $g$, if $E[g(W)|Z=z,A=a,X]= 0$ for almost all $z,a$, then $g(W) = 0$.
\end{assumption}
Assumption~\ref{assump:positivity} is a regular positivity assumption ensuring that in all strata of $X$, there are always some individuals with $A=a,Z=z$ for all $a$, $z$.
Assumption~\ref{assump:completeness} is a commonly used completeness condition for identification \cite{newey2003instrumental}.
Specifically, Assumption~\ref{assump:completeness}(a) essentially requires $U$-comparability. That is, both $Z$ and $W$ should be  associated with $U$ such that variation in $U$ can be recovered from variation in $Z$ and $W$. Assumption~\ref{assump:completeness}(b) aims to ensure that the underlying unmeasured confounding mechanism in $E[Y\mid A,U]$ can be identified using $Z$ and $W$.  	For example, suppose $U$ is a binary variable. Then Assumption~\ref{assump:completeness} further requires that $Z$ and $W$ have at least two categories, and $E[W\mid A=a,Z=1,X=x]-E[W\mid A=a,Z=0,X=x]$ is not equal to zero for all $a,x$.

\paragraph{Rationale}
In the presence of unmeasured confounding by a latent variable $U$, an observed difference in the outcome between the treatment and control groups is a combination  of the underlying causal effect and confounding bias. One cannot directly disentangle the variation in the outcome due to the treatment from the unwanted variation due to $U$, as $U$ is not measured. We seek to indirectly remove such unwanted variation, i.e., unmeasured confounding bias, by leveraging available proxies of $U$. An important example of such proxy is an NCO chosen to be associated with $U$ but not causally affected by the treatment (Figure~\ref{fig:DAG}). Therefore, any difference in the NCO, $W$, between the treatment and control groups can only be attributed to $U$. Such a difference can uncover the unwanted variation due to $U$ assuming that $U$-$Y$ and $U$-$W$ associations are the same, and there is no $U$-$A$ additive interaction on $Y$. An example of such $W$ is the pre-exposure baseline measure of the outcome, in which case bias adjustment reduces to the well-known difference-in-differences approach \cite{sofer2016negative}.

The above describes identification of the ATE under assumptions that are generally untenable, because the $U$-$Y$ and $U$-$W$ associations will often be on different scales, and there may be $U$-$A$ interactions in the model for $Y$. In order to nonparametrically identify unmeasured confounding bias, we make use of the NCE Z. Because $Z$ is associated with $Y$ or $W$ only through $U$, the ratio of $Z$-$Y$ and $Z$-$W$ associations captures the ratio of $U$-$Y$ and $U$-$W$ associations, allowing for $U$-$A$ interactions. 
In summary, leveraging a double negative control design one can nonparametrically identify the magnitude of unmeasured confounding bias via the following mechanism: The NCO uncovers the confounding bias up to a scale that reflects the difference between $U$-$Y$ and $U$-$W$ associations, while the NCE recovers the scale leveraging $Z$-$Y$ and $Z$-$W$ associations. This mechanism is further illustrated in an example below.

\paragraph{Example}
To further illustrate the idea of identification using double negative control, consider a simple example where we assume the following linear structural equation models involving unmeasured confounding $U$, although the nonparametric identification proposed in \cite{miao2018identifying} does not rely on any restriction about the data generating models. We suppress  measured confounders $X$ to ease notation -- all arguments are made implicitly conditional on $X$.

Had $U$ been measured, we could fit (\ref{eq:yau}) and obtain the true causal effect which is $\beta_{\scriptscriptstyle \text{YA}}$. When in fact $U$ is not measured, to leverage double negative control, we additionally assume the $U$-$W$ relationship in (\ref{eq:wu}) and $U$-$Z$ relationship in (\ref{eq:uaz}).
\begin{align}
E[Y\mid A,U] &= \addalpha  \beta_{\scriptscriptstyle \text{YA}}A+\beta_{\scriptscriptstyle \text{YU}}U\label{eq:yau}\\  
E[W\mid U] &= \addbeta \beta_{\scriptscriptstyle \text{WU}}U \label{eq:wu} \\ 
E[U\mid A,Z] &= \addgamma \beta_{\scriptscriptstyle \text{UA}}A+\beta_{\scriptscriptstyle \text{UZ}}Z \label{eq:uaz}.
\end{align}
Models (\ref{eq:yau})--(\ref{eq:uaz}) indicate the following models that one could actually fit using the observed data $(Y,A,W,Z)$. These models are obtained by replacing $U$ with $E[U\mid A,Z]$ in the primary and negative control outcome models (\ref{eq:yau}) and (\ref{eq:wu}).
\begin{alignat}{2}
E[Y\mid A,Z] &\stackrel{(\ref{eq:yau})}{=} \addalpha \beta_{\scriptscriptstyle \text{YA}}A+&&\beta_{\scriptscriptstyle \text{YU}}E[U\mid A,Z]\\ 
&\stackrel{(\ref{eq:uaz})}{=} \addalpha \beta_{\scriptscriptstyle \text{YA}}A+&&\beta_{\scriptscriptstyle \text{YU}}(\addgamma \beta_{\scriptscriptstyle \text{UA}}A+\beta_{\scriptscriptstyle \text{UZ}}Z)\label{eq:yaz}\\   
E[W\mid A,Z]& \stackrel{(\ref{eq:wu})}{=}\addbeta  &&\beta_{\scriptscriptstyle \text{WU}}E[U\mid A,Z] \\
& \stackrel{(\ref{eq:uaz})}{=}\addbeta  &&\beta_{\scriptscriptstyle \text{WU}}(\addgamma \beta_{\scriptscriptstyle \text{UA}}A+\beta_{\scriptscriptstyle \text{UZ}}Z) \label{eq:waz}.
\end{alignat}

From (\ref{eq:yau}) we know that the true causal effect is $\beta_{\scriptscriptstyle \text{YA}}$.
However, if one were to regress  $Y$ on $A$ and $Z$ without accounting for $U$ such as in \cite{flanders2017new}, then the coefficient of $A$ would be equal to $\beta_{\scriptscriptstyle \text{YA}}+\beta_{\scriptscriptstyle \text{YU}}\beta_{\scriptscriptstyle \text{UA}}$. Here $\beta_{\scriptscriptstyle \text{YU}}\beta_{\scriptscriptstyle \text{UA}}$ is confounding bias, which arises when there exists a $U$ that is associated with both $Y$ and $A$. One cannot directly separate the confounding bias from the true causal effect because $U$ is not observed. Nevertheless, the coefficients in the observed models (\ref{eq:yaz}) and (\ref{eq:waz}) allows us to infer $\beta_{\scriptscriptstyle \text{YU}}\beta_{\scriptscriptstyle \text{UA}}$. 
To facilitate discussion, we introduce notation for the coefficients in models (\ref{eq:yaz}) and (\ref{eq:waz}). Let $\delta_A^Y= \beta_{\scriptscriptstyle \text{YA}}+\beta_{\scriptscriptstyle \text{YU}}\beta_{\scriptscriptstyle \text{UA}}$ and $\delta_Z^Y=\beta_{\scriptscriptstyle \text{YU}}\beta_{\scriptscriptstyle \text{UZ}}$ denote the coefficients of $A$ and $Z$ in the primary outcome model (\ref{eq:yaz}), respectively, and let $\delta_A^W=\beta_{\scriptscriptstyle \text{WU}}\beta_{\scriptscriptstyle \text{UA}}$ and $\delta_Z^W=\beta_{\scriptscriptstyle \text{WU}}\beta_{\scriptscriptstyle \text{UZ}}$ denote the coefficients of $A$ and $Z$ in the negative control outcome model (\ref{eq:waz}), respectively.

We detail three strategies to identify the unmeasured confounding bias $\beta_{\scriptscriptstyle \text{YU}}\beta_{\scriptscriptstyle \text{UA}}$ leveraging a single NCO, a single NCE, or the double negative control. First, we note that  coefficient of $A$ in the primary outcome model, $\delta_A^Y$, is a combination of both true causal effect and confounding bias, whereas  coefficient of $A$ in the negative control outcome model, $\delta_A^W$, reflects pure confounding bias because $A$ does not causally affect $W$. In fact, if $U$-$Y$ and $U$-$W$ associations are equal on the additive scale, i.e., $\beta_{\scriptscriptstyle \text{WU}}=\beta_{\scriptscriptstyle \text{YU}}$, then  $\delta_A^W$ matches the confounding bias $\beta_{\scriptscriptstyle \text{YU}}\beta_{\scriptscriptstyle \text{UA}}$. That is, under the assumption of equal $U$-$Y$ and $U$-$W$ additive association, a form of ``additive outcome equi-confounding" \cite{sofer2016negative}, the treatment effect on NCO is equal to the unmeasured confounding bias. Hence the causal effect can be recovered by backing out the association of the treatment with the NCO from the association of the treatment with the primary outcome. Note that in this scenario it is not necessary to have an NCE: one can fit the primary and negative control outcome on treatment without adjusting for the NCE, and then take the difference in treatment effects. When NCO is the baseline outcome, the above reduces to the difference-in-difference method \cite{sofer2016negative}. 
	
	Second, the coefficient of $Z$ in the primary outcome model, $\delta_Z^Y$, would be zero if there was no unmeasured confounding because $Z$ does not causally affect $Y$. Therefore, coefficient of $Z$ in the outcome model reflects pure confounding bias. In fact, if $U$-$A$ and $U$-$Z$ associations are the equal on the additive scale, i.e., $\beta_{\scriptscriptstyle \text{UA}}=\beta_{\scriptscriptstyle \text{UZ}}$, then  $\delta_Z^Y$ captures the bias $\beta_{\scriptscriptstyle \text{YU}}\beta_{\scriptscriptstyle \text{UA}}$ due to unmeasured confounding. That is, under the assumption of equal $U$-$A$ and $U$-$Z$ additive association, a form of ``additive treatment equi-confounding", the NCE effect on the primary outcome is equal to the unmeasured confounding bias. Hence the causal effect is given by the difference in coefficients of treatment and NCE in the primary outcome model. Note that in this scenario it is not necessary to have an NCO: one can fit the primary  outcome on treatment and NCE, and then take the difference in effects of treatment and NCE on $Y$.
	
	In both scenarios described above, the ``additive outcome equi-confounding" or ``additive treatment equi-confounding" is a rather strong assumption, as it requires $Y$ and $W$, or $Z$ and $A$, to operate on the same scale. 
	To relax these assumptions, we can leverage the double negative control. Specifically, if $U$-$Y$ and $U$-$W$ associations are unequal, then $\delta_A^W$  reflects pure confounding bias up to a scale which is equal to $\beta_{\scriptscriptstyle \text{YU}}/\beta_{\scriptscriptstyle \text{WU}}$. Because $Z$-$Y$ ($Z$-$W$) association is a product of $U$-$Z$ and $U$-$Y$ ($U$-$W$) associations, the ratio of $Z$-$Y$ and $Z$-$W$ associations is equal to the ratio of $U$-$Y$ and $U$-$W$ associations. That is, $\beta_{\scriptscriptstyle \text{YU}}/\beta_{\scriptscriptstyle \text{WU}}=\delta_Z^Y/\delta_Z^W$. The confounding bias is thus equal to $\delta_A^W$ scaled by $\delta_Z^Y/\delta_Z^W$, and the true causal effect is give by $\delta_A^Y-\delta_A^W\times \delta_Z^Y/\delta_Z^W$. It is important to note that the first two adjustment methods are a special case of the general adjustment method, in that the confounding bias is always equal to $\delta_A^W\delta_Z^Y/\delta_Z^W$ across all three scenarios.

To summarize, the confounding bias
\begin{subequations}
	\begin{empheq}[left={\beta_{\scriptscriptstyle \text{YU}}\beta_{\scriptscriptstyle \text{UA}}=\delta_A^W \delta_Z^Y/\delta_Z^W= }\empheqlbrace]{align}
\delta_A^W & \text{ if } \beta_{\scriptscriptstyle \text{WU}}=\beta_{\scriptscriptstyle \text{YU}}  \label{eq:deltaaw}  \\
\delta_Z^Y& \text{ if } \beta_{\scriptscriptstyle \text{UA}}=\beta_{\scriptscriptstyle \text{UZ}} \label{eq:deltazy}  \\
\delta_A^W \delta_Z^Y/\delta_Z^W & \text{ if } \beta_{\scriptscriptstyle \text{WU}}\neq \beta_{\scriptscriptstyle \text{YU}} \text{ and } \beta_{\scriptscriptstyle \text{UA}}\neq \beta_{\scriptscriptstyle \text{UZ}} \label{eq:deltaawzy}.
	\end{empheq}
\end{subequations}
Hence the true causal effect is identified as
\begin{equation}\label{eq:identify}
\beta_{\scriptscriptstyle \text{YA}}=\delta_A^Y- \delta_A^W\delta_Z^Y/\delta_Z^W.
\end{equation}
It is important to note that equation (\ref{eq:identify}) is only meaningful when  $\delta_Z^W$ is not equal to zero. If $\delta_Z^W=0$ then either there is no evidence of the presence of $U$ and $\beta_{\scriptscriptstyle \text{YU}}\beta_{\scriptscriptstyle \text{UA}}=0$, or a selected negative control variable is not sufficiently associated with $U$, violating Assumption~\ref{assump:completeness}. Similar arguments apply to $\delta_A^W$ and $\delta_Z^Y$.
In fact, as summarized in Table~\ref{tbl:method}, many negative control methods detect, reduce, and remove unmeasured confounding bias using analogies of scenario (\ref{eq:deltaaw}) \cite{richardson2015negative,richardson2014assessment,tchetgen2015negative,sofer2016negative} and scenario (\ref{eq:deltazy}) \cite{flanders2011method,flanders2017new,miao2017invited}.

In practice, identification via (\ref{eq:identify}) relies on fitting the primary and negative control outcome models $E[Y\mid A,Z]$ and $E[W\mid A,Z]$. Alternatively, one could directly make assumption about the underlying unmeasured confounding mechanism $E[Y\mid A,U]$ which is proposed in \cite{miao2018confounding}. To illustrate, consider again the example above. Let $\widetilde{U}_{W}=\frac{W-\beta_{W0}}{\beta_{\scriptscriptstyle \text{WU}}}$, then by (\ref{eq:wu}) $\widetilde{U}_{W}$ is a good proxy of $U$ in the sense that $E[\widetilde{U}_{W}\mid U]=U$.
In particular, let $h(W,A) = \addalpha \beta_{\scriptscriptstyle \text{YA}}A+\beta_{\scriptscriptstyle \text{YU}}\widetilde{U}_{W}$, then by (\ref{eq:yau}) we have   
\begin{align}
E[Y\mid A,U]&=E[h(W,A)\mid A,U], \label{eq:hau}\\
E[Y\mid A,Z]&=E[h(W,A)\mid A,Z], \label{eq:haz}
\end{align}
where (\ref{eq:haz}) is obtained by taking expectation on both sides of (\ref{eq:hau}).
The above equations indicate that  $h$ captures  the relationship between $U$-$Y$ and $U$-$W$ associations via (\ref{eq:hau}), which can be identified by the relationship between $Z$-$Y$ and $Z$-$W$ associations via (\ref{eq:haz}). Because of this key observation, $h$ is referred to as the confounding bridge function in \cite{miao2018confounding}. {The functional form of $h$ is implied by (\ref{eq:yau}) and (\ref{eq:wu}).} Once $h$ is identified, we have that  $E[Y(a)]\stackrel{(\ref{eq:hau}) }{=}E_{U}\{E[Y\mid A=a,U]\}=E[h(W,A=a)]$.
In practice, one  may assume a familiar linear model about the functional  form of $h$ that satisfies (\ref{eq:hau}), such as
\begin{equation}
h(W,A;\theta)=\theta_0+\theta_AA+\theta_WW.
\end{equation}
Then under Assumption~\ref{assump:completeness}, $\theta$ can be identified by the population moment equation  $E[g(A,Z)\{Y-h(W,A;\theta)\}]=0$ using the generalized method of moments (GMM) method \cite{hansen1982large}.
With $\theta$ identified, the ATE is given by 
\begin{equation}
\text{ATE} =E[h(W,A=1;\theta)]-E[h(W,A=0;\theta)].
\end{equation}
A simple version of the above GMM procedure can be realized via a simple two stage least squares procedure as followed \cite{miao2018confounding}:
\begin{align*}
&\text{Stage I: regress $W$ on $A$ and $Z$ (with intercept), {and obtain the fitted value $\widehat{W}$ as a proxy of $U$};}\\
&\text{Stage II: regress $Y$ on $A$ (with intercept), adjusting for $\widehat{W}$,}
\end{align*}
 then the coefficient of $A$ is the true causal effect $\beta_{\scriptscriptstyle \text{YA}}$ assuming (\ref{eq:yau}) and (\ref{eq:wu}). The two stage least squares approach given above provides a simple implementation of the NC method using existing and widely disseminated IV software packages such as the \texttt{ivregress}, \texttt{ivreg}, or \texttt{ivreg2} command in Stata, the \texttt{gmm}, \texttt{sem}, \texttt{ivpack}, or \texttt{AER} package in R, and the \texttt{SYSLIN} procedure in SAS. 

\section{Conclusions\label{sec:discussion}} 
Negative controls are innovative and important tools in observational studies. Development of negative control methods will encourage researchers to routinely check for evidence of confounding bias and rigorously adjust for residual confounding bias. Negative control variables are widely available in routinely collected healthcare data such as administrative claims and electronic health records data, because information on secondary treatments and outcomes beyond the primary treatment and outcome of interest are often recorded, and such secondary treatments and outcomes can potentially serve as negative controls. Therefore development of negative controls methods is critical to unlocking the full potential of contemporary healthcare data and ultimately improve the validity of research findings. It is important to note that other sources of bias, such as selection bias and misclassification bias, are typical in routinely collected healthcare data.  Developing negative control methods accounting for bias beyond residual confounding is thus an important area of future research.

We have specified statistical assumptions, practical strategies, and validation criteria that can be combined with subject matter knowledge to design negative control studies in Section~\ref{sec:application}. We also illustrated identification of the ATE by either fitting the observed primary and negative control outcome models or through assumption on the unmeasured confounding mechanism followed by a simple two stage least squares procedure  in Section~\ref{sec:method}. We believe that these examples can provide practical guidance on use of negative control methods to a broader audience.

\clearpage
\appendix
\renewcommand\thesubsection{A.\arabic{subsection}}
\renewcommand\thetable{A.\arabic{table}}
\setcounter{table}{0}
\section*{Appendix}
\subsection{Examples of invalid negative controls that violates some assumption}
\paragraph*{Violation 1: no arrow between U and W}
There must be an arrow between $U$ and $W$, because an NCO is a proxy of unmeasured confounder. It recovers the confounding bias by reflecting variation due to $U$. 

\paragraph*{Violation 2: no arrow between U and Z, and Z$\not\to$A}
The only scenario that $Z$ does not need to be associated with $U$ is when $Z$ is an instrumental variable (see first cell of Table~\ref{example_ZW}). In this case, $A$ is a collider between $Z$ and $U$, such that $Z$ and $U$ are marginally independent.  Conditioning on a collider will create collider bias such that $Z$ and $U$ become conditionally dependent. The requirements about $Z$ in Assumptions~\ref{assump:ucompare} and \ref{assump:completeness} are all made conditioning on $A$. Therefore an instrumental variable is a valid NCE.

\paragraph*{Violation 3: Y$\to$W} If the outcome causes the NCO, then the treatment directly causes the NCO via the path $A\rightarrow Y\rightarrow W$, which violates Assumption~\ref{assump:nco}.

\paragraph*{Violation 4: Z$\rightarrow$U$\leftarrow$W} The direction of the arrow between $U$ and the negative control doesn't always matter. For example, we can have $Z\to U$, $U\to Z$, $W\to U$, or $U\to W$. However, if both $Z$ and $W$ cause $U$, then $U$ is a collider in the path $Z\rightarrow U\leftarrow W$. In this case, conditional on $U$, $Z$ and $W$ will become associated. This violates Assumption~\ref{assump:nce}. 

\subsection{Example of causal graphs encoding the negative control assumptions}
Below  we enumerate the possible relationships among $Z,A,U$  and among $Y,W,U$ in Table~\ref{example_ZW}. 
These partial graphs can be combined into a directed acyclic graph that encodes the negative control assumptions. Grey colored graphs are invalid because of violation of key assumptions.

\clearpage
\newgeometry{margin=1cm}
\begin{landscape}
\begin{table}[h]
	\caption{Examples of graphs for $Z,A,U$ relationships and for $W,Y,U$ relationships. The two pieces of graphs can be combined in to a  directed acyclic graph that encodes the negative control assumptions. Grey colored graphs are invalid because of violation of key assumptions.}\label{example_ZW}
	\centering{\small
	\renewcommand{\arraystretch}{1.45}
	\begin{tabular}{|c|c|c|c|}
		\hline 
		\multicolumn{4}{|c|}{Examples of graphs for $Z,A,U$ relationships}\tabularnewline
		\hline 
		& $Z\rightarrow A$ (pre-treatment) & $A\rightarrow Z$  (post-treatment) & $Z\ind A$ \tabularnewline
		\hline 
		No arrow between & Instrumental variable (IV) & Violate Assumption~\ref{assump:ucompare} and \ref{assump:completeness} & Violate Assumption~\ref{assump:ucompare} and \ref{assump:completeness}  \tabularnewline
		 $U$ and $Z$ (may violate & \multirow{2}{*}{\resizebox{1.25in}{0.4in}{
				\begin{tikzpicture}
				\tikzset{line width=1pt,inner sep=5pt,
					ell/.style={draw, inner sep=1.5pt,line width=1pt}}
				\node[shape=ellipse,ell] (A) at (1,-1.5) {$A$};
				\node[shape=ellipse,ell] (U) at (2,-0.5) {$U,X$};
				\node[shape=ellipse,ell] (Y) at (3,-1.5) {$Y$};
				\node[shape=circle,ell,inner sep=2.2pt] (Z) at (-1.1,-1.5) {$Z$};
				\foreach \from/\to in {U/A,U/Y,A/Y,Z/A}
				\draw[-stealth,line width=0.5pt] (\from) -- (\to);
				\end{tikzpicture}} } & \multirow{2}{*}{\resizebox{1.25in}{0.4in}{
				\begin{tikzpicture}
				\tikzset{line width=1pt,inner sep=5pt, color=black!30,
					ell/.style={draw, inner sep=1.5pt,line width=1pt,color=black!30}}
				\node[shape=ellipse,ell] (A) at (1,-1.5) {$A$};
				\node[shape=ellipse,ell] (U) at (2,-0.5) {$U,X$};
				\node[shape=ellipse,ell] (Y) at (3,-1.5) {$Y$};
				\node[shape=circle,ell,inner sep=2.2pt] (Z) at (-1.1,-1.5) {$Z$};
				\foreach \from/\to in {U/A,U/Y,A/Y,A/Z}
				\draw[-stealth,line width=0.5pt,color=black!30] (\from) -- (\to);
				\end{tikzpicture}} } & \multirow{2}{*}{\resizebox{1.25in}{0.4in}{
				\begin{tikzpicture}
				\tikzset{line width=1pt,inner sep=5pt, color=black!30,
					ell/.style={draw, inner sep=1.5pt,line width=1pt,color=black!30}}
				\node[shape=ellipse,ell] (A) at (1,-1.5) {$A$};
				\node[shape=ellipse,ell] (U) at (2,-0.5) {$U,X$};
				\node[shape=ellipse,ell] (Y) at (3,-1.5) {$Y$};
				\node[shape=circle,ell,inner sep=2.2pt] (Z) at (-1.1,-1.5) {$Z$};
				\foreach \from/\to in {U/A,U/Y,A/Y}
				\draw[-stealth,line width=0.5pt,color=black!30] (\from) -- (\to);
				\end{tikzpicture}} } \tabularnewline
		Assumption~\ref{assump:ucompare} and \ref{assump:completeness})& & &  \tabularnewline
		\cline{1-4} 
		& Invalid IV & Post-treatment proxy of $U$ & Surrogate of $U$ \tabularnewline
		$U\rightarrow Z$ & \multirow{2}{*}{\resizebox{1.25in}{0.4in}{
				\begin{tikzpicture}
				\tikzset{line width=1pt,inner sep=5pt, 
					ell/.style={draw, inner sep=1.5pt,line width=1pt}}
				\node[shape=ellipse,ell] (A) at (1,-1.5) {$A$};
				\node[shape=ellipse,ell] (U) at (2,-0.5) {$U,X$};
				\node[shape=ellipse,ell] (Y) at (3,-1.5) {$Y$};
				\node[shape=circle,ell,inner sep=2.2pt] (Z) at (-1.1,-1.5) {$Z$};
				\foreach \from/\to in {U/A,U/Y,A/Y,U/Z,Z/A}
				\draw[-stealth,line width=0.5pt] (\from) -- (\to);
				\end{tikzpicture}}} & \multirow{2}{*}{\resizebox{1.25in}{0.4in}{
				\begin{tikzpicture}
				\tikzset{line width=1pt,inner sep=5pt,
					ell/.style={draw, inner sep=1.5pt,line width=1pt}}
				\node[shape=ellipse,ell] (A) at (1,-1.5) {$A$};
				\node[shape=ellipse,ell] (U) at (2,-0.5) {$U,X$};
				\node[shape=ellipse,ell] (Y) at (3,-1.5) {$Y$};
				\node[shape=circle,ell,inner sep=2.2pt] (Z) at (-1.1,-1.5) {$Z$};
				\foreach \from/\to in {U/A,U/Y,A/Y,U/Z,A/Z}
				\draw[-stealth,line width=0.5pt] (\from) -- (\to);
				\end{tikzpicture}}} & \multirow{2}{*}{\resizebox{1.25in}{0.4in}{
				\begin{tikzpicture}
				\tikzset{line width=1pt,inner sep=5pt,
					ell/.style={draw, inner sep=1.5pt,line width=1pt}}
				\node[shape=ellipse,ell] (A) at (1,-1.5) {$A$};
				\node[shape=ellipse,ell] (U) at (2,-0.5) {$U,X$};
				\node[shape=ellipse,ell] (Y) at (3,-1.5) {$Y$};
				\node[shape=circle,ell,inner sep=2.2pt] (Z) at (-1.1,-1.5) {$Z$};
				\foreach \from/\to in {U/A,U/Y,A/Y,U/Z}
				\draw[-stealth,line width=0.5pt] (\from) -- (\to);
				\end{tikzpicture}}} \tabularnewline
		& & &  \tabularnewline
		\hline 
		& \multicolumn{3}{c|}{May violate Assumption~\ref{assump:nce} if there is $W\rightarrow U$} \tabularnewline
		$Z\rightarrow U$ & \multirow{2}{*}{\resizebox{1.25in}{0.4in}{
				\begin{tikzpicture}
				\tikzset{line width=1pt,inner sep=5p,
					ell/.style={draw, inner sep=1.5pt,line width=1pt}}
				\node[shape=ellipse,ell] (A) at (1,-1.5) {$A$};
				\node[shape=ellipse,ell] (U) at (2,-0.5) {$U,X$};
				\node[shape=ellipse,ell] (Y) at (3,-1.5) {$Y$};
				\node[shape=circle,ell,inner sep=2.2pt] (Z) at (-1.1,-1.5) {$Z$};
				\foreach \from/\to in {U/A,U/Y,A/Y,Z/U,Z/A}
				\draw[-stealth,line width=0.5pt] (\from) -- (\to);
				\end{tikzpicture}}} & \multirow{2}{*}{\resizebox{1.25in}{0.4in}{
				\begin{tikzpicture}
				\tikzset{line width=1pt,inner sep=5pt, color=black!30,
					ell/.style={draw, inner sep=1.5pt,line width=1pt, color=black!30}}
				\node[shape=ellipse,ell] (A) at (1,-1.5) {$A$};
				\node[shape=ellipse,ell] (U) at (2,-0.5) {$U,X$};
				\node[shape=ellipse,ell] (Y) at (3,-1.5) {$Y$};
				\node[shape=circle,ell,inner sep=2.2pt] (Z) at (-1.1,-1.5) {$Z$};
				\foreach \from/\to in {U/A,U/Y,A/Y,Z/U,A/Z}
				\draw[-stealth,line width=0.5pt, color=black!30] (\from) -- (\to);
				\end{tikzpicture}}} & \multirow{2}{*}{\resizebox{1.25in}{0.4in}{
				\begin{tikzpicture}
				\tikzset{line width=1pt,inner sep=5pt,
					ell/.style={draw, inner sep=1.5pt,line width=1pt}}
				\node[shape=ellipse,ell] (A) at (1,-1.5) {$A$};
				\node[shape=ellipse,ell] (U) at (2,-0.5) {$U,X$};
				\node[shape=ellipse,ell] (Y) at (3,-1.5) {$Y$};
				\node[shape=circle,ell,inner sep=2.2pt] (Z) at (-1.1,-1.5) {$Z$};
				\foreach \from/\to in {U/A,U/Y,A/Y,Z/U}
				\draw[-stealth,line width=0.5pt] (\from) -- (\to);
				\end{tikzpicture}}} \tabularnewline
		& & & \tabularnewline
		\hline 
		\multicolumn{4}{|c|}{Examples of graphs for $W,Y,U$ relationships}\tabularnewline
		\hline 
		& $W\rightarrow Y(a)$ & $Y(a)\rightarrow W$ & $Y(a)\ind W\mid(U,X)$\tabularnewline
		& & (violate Assumptions~\ref{assump:nco} and \ref{assump:nce}) & \tabularnewline
		\hline 
		No arrow between& Violate Assumption~\ref{assump:ucompare} and \ref{assump:completeness} & Violate Assumptions~\ref{assump:nco}, \ref{assump:ucompare}, and \ref{assump:completeness} & Violate Assumption~\ref{assump:ucompare} and \ref{assump:completeness}\tabularnewline
		$U$ and $W$ (violate & \multirow{2}{*}{\resizebox{1.25in}{0.4in}{
				\begin{tikzpicture}
				\tikzset{line width=1pt,inner sep=5pt, color=black!30,
					ell/.style={draw, inner sep=1.5pt,line width=1pt,color=black!30}}
				\node[shape=ellipse,ell] (A) at (1,-1.5) {$A$};
				\node[shape=ellipse,ell] (U) at (2,-0.5) {$U,X$};
				\node[shape=ellipse,ell] (Y) at (3,-1.5) {$Y$};
				\node[shape=circle,ell] (W) at (5,-1.5) {$W$};
				\foreach \from/\to in {U/A,U/Y,A/Y,W/Y}
				\draw[-stealth,line width=0.5pt,color=black!30] (\from) -- (\to);
				\end{tikzpicture}}} & \multirow{2}{*}{\resizebox{1.25in}{0.4in}{
				\begin{tikzpicture}
				\tikzset{line width=1pt,inner sep=5pt, color=black!30,
					ell/.style={draw, inner sep=1.5pt,line width=1pt,color=black!30}}
				\node[shape=ellipse,ell] (A) at (1,-1.5) {$A$};
				\node[shape=ellipse,ell] (U) at (2,-0.5) {$U,X$};
				\node[shape=ellipse,ell] (Y) at (3,-1.5) {$Y$};
				\node[shape=circle,ell] (W) at (5,-1.5) {$W$};
				\foreach \from/\to in {U/A,U/Y,A/Y,Y/W}
				\draw[-stealth,line width=0.5pt,color=black!30] (\from) -- (\to);
				\end{tikzpicture}}} & \multirow{2}{*}{\resizebox{1.25in}{0.4in}{
				\begin{tikzpicture}
				\tikzset{line width=1pt,inner sep=5pt, color=black!30,
					ell/.style={draw, inner sep=1.5pt,line width=1pt,color=black!30}}
				\node[shape=ellipse,ell] (A) at (1,-1.5) {$A$};
				\node[shape=ellipse,ell] (U) at (2,-0.5) {$U,X$};
				\node[shape=ellipse,ell] (Y) at (3,-1.5) {$Y$};
				\node[shape=circle,ell] (W) at (5,-1.5) {$W$};
				\foreach \from/\to in {U/A,U/Y,A/Y}
				\draw[-stealth,line width=0.5pt,color=black!30] (\from) -- (\to);
				\end{tikzpicture}}}\tabularnewline
		Assumption~\ref{assump:ucompare} and \ref{assump:completeness})& & & \tabularnewline
		\hline 
		& & Violate Assumption~\ref{assump:nco} & \tabularnewline
		$U\rightarrow W$ & \multirow{2}{*}{\resizebox{1.25in}{0.4in}{
				\begin{tikzpicture}
				\tikzset{line width=1pt,inner sep=5pt,
					ell/.style={draw, inner sep=1.5pt,line width=1pt}}
				\node[shape=ellipse,ell] (A) at (1,-1.5) {$A$};
				\node[shape=ellipse,ell] (U) at (2,-0.5) {$U,X$};
				\node[shape=ellipse,ell] (Y) at (3,-1.5) {$Y$};
				\node[shape=circle,ell] (W) at (5,-1.5) {$W$};
				\foreach \from/\to in {U/A,U/Y,A/Y,W/Y,U/W}
				\draw[-stealth,line width=0.5pt] (\from) -- (\to);
				\end{tikzpicture}}} & \multirow{2}{*}{\resizebox{1.25in}{0.4in}{
				\begin{tikzpicture}
				\tikzset{line width=1pt,inner sep=5pt, color=black!30,
					ell/.style={draw, inner sep=1.5pt,line width=1pt,color=black!30}}
				\node[shape=ellipse,ell] (A) at (1,-1.5) {$A$};
				\node[shape=ellipse,ell] (U) at (2,-0.5) {$U,X$};
				\node[shape=ellipse,ell] (Y) at (3,-1.5) {$Y$};
				\node[shape=circle,ell] (W) at (5,-1.5) {$W$};
				\foreach \from/\to in {U/A,U/Y,A/Y,Y/W,U/W}
				\draw[-stealth,line width=0.5pt,color=black!30] (\from) -- (\to);
				\end{tikzpicture}}} & \multirow{2}{*}{\resizebox{1.25in}{0.4in}{
				\begin{tikzpicture}
				\tikzset{line width=1pt,inner sep=5pt,
					ell/.style={draw, inner sep=1.5pt,line width=1pt}}
				\node[shape=ellipse,ell] (A) at (1,-1.5) {$A$};
				\node[shape=ellipse,ell] (U) at (2,-0.5) {$U,X$};
				\node[shape=ellipse,ell] (Y) at (3,-1.5) {$Y$};
				\node[shape=circle,ell] (W) at (5,-1.5) {$W$};
				\foreach \from/\to in {U/A,U/Y,A/Y,U/W}
				\draw[-stealth,line width=0.5pt] (\from) -- (\to);
				\end{tikzpicture}}}\tabularnewline
		& & & \tabularnewline
		\hline 
		& \multicolumn{3}{c|}{May violate Assumption~\ref{assump:nce} if there is $Z\rightarrow U$}\tabularnewline
		& & Violate Assumption~\ref{assump:nco} & \tabularnewline
		$W\rightarrow U$ & \multirow{2}{*}{\resizebox{1.25in}{0.4in}{
				\begin{tikzpicture}
				\tikzset{line width=1pt,inner sep=5pt,
					ell/.style={draw, inner sep=1.5pt,line width=1pt}}
				\node[shape=ellipse,ell] (A) at (1,-1.5) {$A$};
				\node[shape=ellipse,ell] (U) at (2,-0.5) {$U,X$};
				\node[shape=ellipse,ell] (Y) at (3,-1.5) {$Y$};
				\node[shape=circle,ell] (W) at (5,-1.5) {$W$};
				\foreach \from/\to in {U/A,U/Y,A/Y,W/Y,W/U}
				\draw[-stealth,line width=0.5pt] (\from) -- (\to);
				\end{tikzpicture}}} & \multirow{2}{*}{\resizebox{1.25in}{0.4in}{
				\begin{tikzpicture}
				\tikzset{line width=1pt,inner sep=5pt, color=black!30,
					ell/.style={draw, inner sep=1.5pt,line width=1pt,color=black!30}}
				\node[shape=ellipse,ell] (A) at (1,-1.5) {$A$};
				\node[shape=ellipse,ell] (U) at (2,-0.5) {$U,X$};
				\node[shape=ellipse,ell] (Y) at (3,-1.5) {$Y$};
				\node[shape=circle,ell] (W) at (5,-1.5) {$W$};
				\foreach \from/\to in {U/A,U/Y,A/Y,Y/W,W/U}
				\draw[-stealth,line width=0.5pt,color=black!30] (\from) -- (\to);
				\end{tikzpicture}}} & \multirow{2}{*}{\resizebox{1.25in}{0.4in}{
				\begin{tikzpicture}
				\tikzset{line width=1pt,inner sep=5pt,
					ell/.style={draw, inner sep=1.5pt,line width=1pt}}
				\node[shape=ellipse,ell] (A) at (1,-1.5) {$A$};
				\node[shape=ellipse,ell] (U) at (2,-0.5) {$U,X$};
				\node[shape=ellipse,ell] (Y) at (3,-1.5) {$Y$};
				\node[shape=circle,ell] (W) at (5,-1.5) {$W$};
				\foreach \from/\to in {U/A,U/Y,A/Y,W/U}
				\draw[-stealth,line width=0.5pt] (\from) -- (\to);
				\end{tikzpicture}}}\tabularnewline
		& & & \tabularnewline
		\hline 
	\end{tabular}
}
\end{table}
\end{landscape}
\restoregeometry

\clearpage
\printbibliography
\end{document}